\documentclass[10pt,two-column]{article}

\usepackage[utf8]{inputenc}
\usepackage[T1]{fontenc}
\usepackage{amsmath,amsfonts,amssymb}
\usepackage{graphicx}
\usepackage{float}
\usepackage{caption}
\usepackage{subcaption}
\usepackage{geometry}
\usepackage{times}
\usepackage{url}
\usepackage[colorlinks=true,allcolors=blue]{hyperref}
\usepackage{booktabs}
\usepackage{multirow}
\usepackage{siunitx}
\usepackage[auth-lg]{authblk}
\usepackage{comment} 

\makeatletter
\renewcommand{\maketitle}{
    \twocolumn[
        \begin{@twocolumnfalse}
        \begin{center}
            {\Large\bfseries \@title \par}
            \vspace{1em}
            {\normalsize \@author \par}
            \vspace{0.5em}
            {\normalsize \@date \par}
            \vspace{1.5em}
        \end{center}
        \end{@twocolumnfalse}
    ]
}
\makeatother

\renewenvironment{abstract}{
	\bfseries
    \small
}{\par\bigskip}

\usepackage{titlesec}
\titleformat{\section}
  {\normalfont\large\bfseries}{\thesection}{1em}{}
\titlespacing{\section}
	{0cm}{1em}{0.1em}
\titleformat{\subsection}
  {\normalfont\normalsize\bfseries}{\thesubsection}{1em}{}
\titlespacing{\subsection}
	{0cm}{0.2em}{0em}

\newcommand{\mm}{Mu}

\usepackage[
  bibstyle=nature,
  natbib=true,
  sorting=none,
  maxnames=1000,
  minnames=1,
  backend=bibtex8,
  doi=true,
  ]{biblatex}
  
\AtEveryBibitem{%
  \clearfield{day}%
  \clearfield{month}%
  \clearfield{endday}%
  \clearfield{endmonth}%
}

\begin{document}

\title{Synthesis of a high intensity, superthermal muonium beam for \\ gravity and laser spectroscopy experiments}

\author[1]{J. Zhang}
\author[1,2]{A. Antognini}
\author[2]{M. Bartkowiak}
\author[1,2]{K. Kirch}
\author[2]{A. Knecht}
\author[1]{D. Goeldi}
\author[1,2]{D. Taqqu}
\author[1]{R. Waddy}
\author[3]{F. Wauters}
\author[1]{P. Wegmann}
\author[1,*]{A. Soter}

\affil[1]{Institute for Particle Physics and Astrophysics, ETH Zürich, 8093 Zürich, Switzerland}
\affil[2]{PSI Center for Neutron and Muon Sciences, 5232 Villigen-PSI, Switzerland}
\affil[3]{Institute for Nuclear Physics, Johannes Gutenberg-University, D-55122 Mainz, Germany}
\affil[*]{Corresponding author: asoter@phys.ethz.ch }

\date{}

\maketitle

\begin{abstract}
The universality of free fall, a cornerstone of Einstein’s theory of gravity \cite{einstein_grundlage_1916}, has so far only been tested with neutral composite states of first-generation Standard Model (SM) particles, such as atoms  \cite{albers_quantum_2020, asenbaum_atominterferometric_2020} or neutrons \cite{schmiedmayer_equivalence_1989} and, most recently, antihydrogen \cite{anderson_observation_2023}. 
Extending these gravitational measurements to other sectors of the SM requires the formation of neutral bound states using higher-generation, unstable particles. Muonium, the bound state of an antimuon ($\mu^+$) and an electron ($e^-$), offers the possibility to probe gravity with second-generation (anti)leptons, in the absence of the strong interaction.  
However, the short $\mu^+$ lifetime ($\tau_{\mu}\approx \SI{2.2}{\us}$) and the existing diffuse thermal muonium sources rendered such measurements unfeasible. Here, we report the synthesis of a high-brightness muonium beam, extracted from a thin layer of superfluid helium by exploiting its chemical potential and unique transport properties. The mean longitudinal velocity (${v}\approx 2180~\rm{m/s}$) and narrow distribution (${\Delta v}< 150 ~\rm{m/s}$) of the atoms characterise a superthermal beam, while yields are similar to the highest intensity diffuse sources. This new beam is expected to enable muonium interferometry and a percent-level measurement of its gravitational acceleration, providing the first direct test of the Weak Equivalence Principle with second-generation (anti)matter. Its unprecedented brightness also opens the way to sub-kHz 1S-2S spectroscopy, enabling precise determination of the muon mass and stringent tests of bound-state quantum electrodynamics.

\end{abstract}


Muonium (\mm) is a hydrogen-like exotic atom, where a positive anti-muon ($\mu^+$) and an electron ($e^-$) are bound by the Coulomb interaction. This purely leptonic system is a unique precision probe to test bound-state quantum electrodynamics (QED) without the influence of nuclear and finite-size effects \cite{karshenboim_precision_2005}. Laser spectroscopy of the {\mm} 1S-2S transition \cite{meyer_measurement_2000}, and microwave spectroscopy of the {\mm} ground state hyperfine structure \cite{liu_high_1999} provide precision measurements of fundamental constants like the muon mass and magnetic moment, while searches for muonium-antimuonium conversion constrain charged lepton number violation \cite{willmann_new_1999}. Future studies and ongoing efforts for improvements \cite{crivelli_mumass_2018, ohayon_precision_2022, strasser2025precision} are strongly motivated by other precision measurements like the anomalous magnetic moment of the muon \cite{themuong2collaboration_measurement_2023, delaunay_independent_2021}, and searches for new physics \cite{stadnik_searching_2023,gomes_laboratory_2014}.

Another unique and so far unexplored facet of {\mm} is that its mass is dominated by the $\mu^+$ mass, which is not only an elementary antiparticle, but also a second-generation lepton. Direct measurements of its gravitational acceleration $g_{\rm\mm}$ would provide a unique way to test the universality of free fall, and hence Einstein's Weak Equivalence Principle (WEP) in higher generations of the Standard Model (SM). The WEP was rigorously tested using normal matter probes in lab-based \cite{,wagner_torsionbalance_2012} and satellite-borne \cite{microscopecollaboration_microscope_2022} experiments to $\sim10^{-15}$ fractional precision. Less precise studies extended the mass scale from single atoms \cite{albers_quantum_2020, asenbaum_atominterferometric_2020} and neutrons \cite{schmiedmayer_equivalence_1989} to celestial objects exerting their own gravitational field \cite{williams_lunar_2012}. Since the relative strength of gravity is $\sim$30-40 orders of magnitude weaker than the electromagnetic interaction between charged particles, neutral exotic atoms are so far the only feasible platforms to directly study gravity in other sectors of the SM. Recently, the ALPHA experiment at CERN confirmed that antihydrogen ($\bar{\rm H}=\bar{p} + e^+$) falls towards Earth with an acceleration consistent with normal matter \cite{anderson_observation_2023}.  Earlier indirect measurements using charged particles in Penning traps  \cite{hughes_constraints_1991, borchert_16partspertrillion_2022}, and neutral meson oscillations \cite{karshenboim_neutral_2009, apostolakis_tests_1999} also implied equivalency between matter and antimatter, and further experiments using antihydrogen \cite{perez_gbar_2015,doser_aegis_2018} and positronium ($\rm{Ps}= e^- + e^+$) \cite{cassidy_atom_2014} have been proposed. 

All gravitational free-fall measurements to date involved, however, composite hadrons: protons, neutrons or antiprotons, which are confined states of first generation ($u,d$) (anti)quarks. Quark masses attributed to the Higgs mechanism constitute only a few percent of the hadron masses, while the rest arises from dynamics induced by the strong interaction \cite{yang_proton_2018}. In \mm{}, by contrast, the electromagnetic binding energy $Q_{\rm em}\approx -13.5~$eV causes negligible defect to the constituent masses ($m_{\rm \mm{}}=m_e+m_{\mu}+Q_{\rm em}\approx 106.17$~MeV/$c^2$), which remains dominated by the muon mass. Hence, a \mm{} free-fall experiment would be the first direct test of gravity acting predominantly on the mass content of the stress-energy tensor. While mainly antimatter (antihydrogen) gravity was discussed on the theoretical side \cite{nieto_arguments_1991,karshenboim_neutral_2009,villata_cpt_2011,benoit-levy_introducing_2012}, the free fall of \mm{} offers to test a flavour-dependent coupling of gravity in the second generation, and to probe beyond-SM interactions between electrons and muons \cite{stadnik_searching_2023}.

\begin{figure*}[ht]
\includegraphics[width=\textwidth]{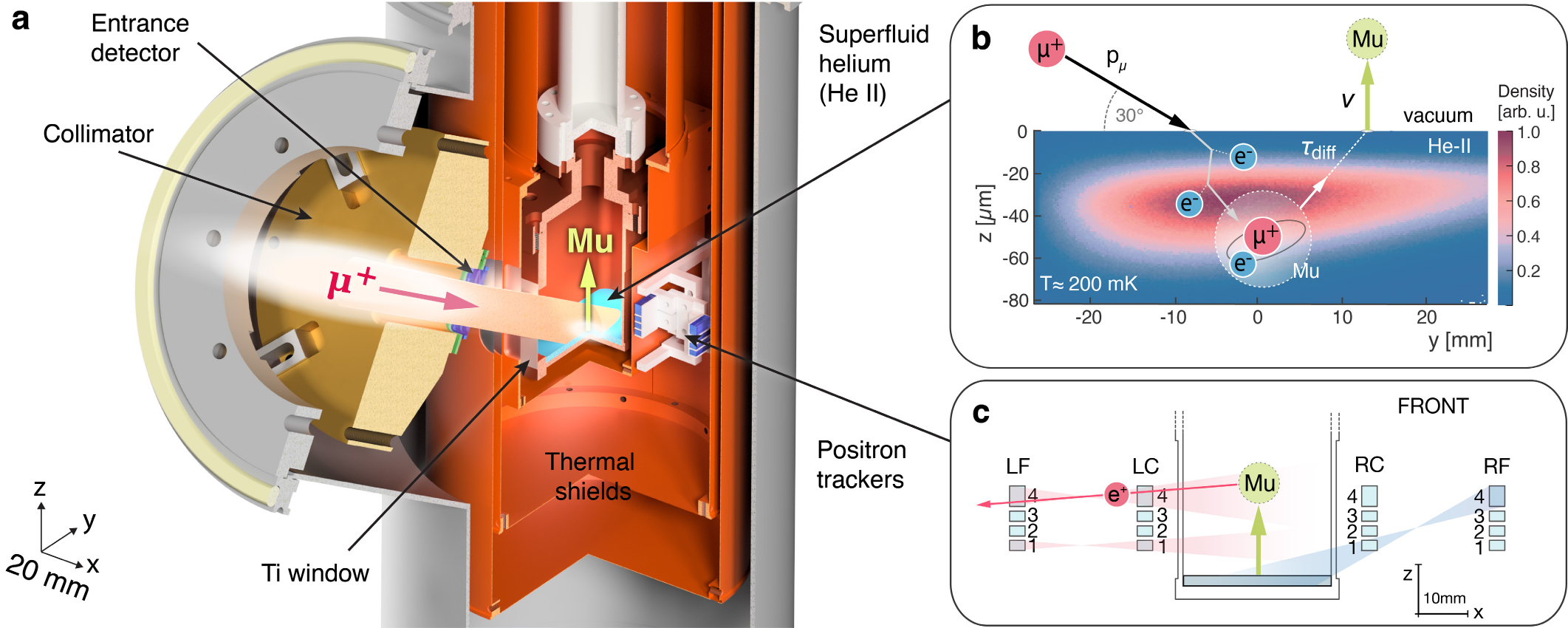}\caption{\label{fig:Setup} \textbf{Experimental principle and layout.} (a) The superfluid helium (He II) chamber and dilution refrigerator. After traversing the entrance detector and several thin foils, the $\mu^+$ comes to rest in a thin horizontal He~II layer. (b) Principle of Mu stopping and formation in He II, with the simulated $\mu^+$ stopping distribution (coloured gradient in background). The incoming $\mu^+$ comes to rest in an average $\approx 35~\mu$m depth, and captures an electron from the ionization trail. The formed Mu diffuses ballistically to the surface and is emitted perpendicularly with a velocity defined by the chemical potential. (c) Schematic of the positron tracking system to localize $\mu^+$ decays. The acceptance areas of selected detector coincidences are sketched, with red shaded areas outlying Layer 1 and Layer 4 coincidences on the left side (LC1$\land$LF1, LC4$\land$LF4), and blue denoting one of the coincidence conditions (RC1$\land$RF4) monitoring the target liquid.}
\end{figure*}

Carrying out this experiment is inherently challenging due to the short lifetime of the $\mu^+$ ($\tau_{\mu} \approx \SI{2.2}{\us}$), and the difficulties to synthesize the atoms in vacuum. Measuring the small influence of gravity in such short timescales requires phase-sensitive methods like Talbot-Lau or Mach-Zehnder interferometry \cite{oberthaler_inertial_1996}, which is the proposed approach of the LEMING experiment \cite{soter_development_2021}. Interferometry requires a \mm{} beam in vacuum, with high intensity and narrow angular- and momentum distribution that conventional \mm{} sources could not provide. Prior to this work, vacuum \mm{} sources were based on porous materials such as SiO$_2$ powder \cite{antognini_roomtemperature_2022} or aerogel \cite{beare_study_2020}, in which muons could stop and recombine with electrons to form \mm{}. Typically a few percent of the formed \mm{} atoms reached the surface within their lifetime via diffusion through the nanoscopic pores \cite{beare_study_2020,antognini_roomtemperature_2022}, which were emitted to vacuum with a thermal (Maxwell-Boltzmann) velocity distribution and broad ($\sim\cos\theta$) angular distribution \cite{zhang_modeling_2022}. Such thermal \mm{} beams would require drastic collimation and velocity selection for interferometry, which would reduce the number of atoms beyond feasibility for a gravity measurement. The low yield and large angular and velocity spreads also introduce statistical and systematic uncertainties for precision \mm{} laser spectroscopy \cite{javary_twophoton_2025}. While cooling these converters narrows the momentum distribution of the \mm{} atoms, the yields drop rapidly with the temperature, with essentially no \mm{} diffusing to vacuum at below 100~K \cite{antognini_muonium_2012}.

\begin{figure}[!ht]
\includegraphics[width=\linewidth]{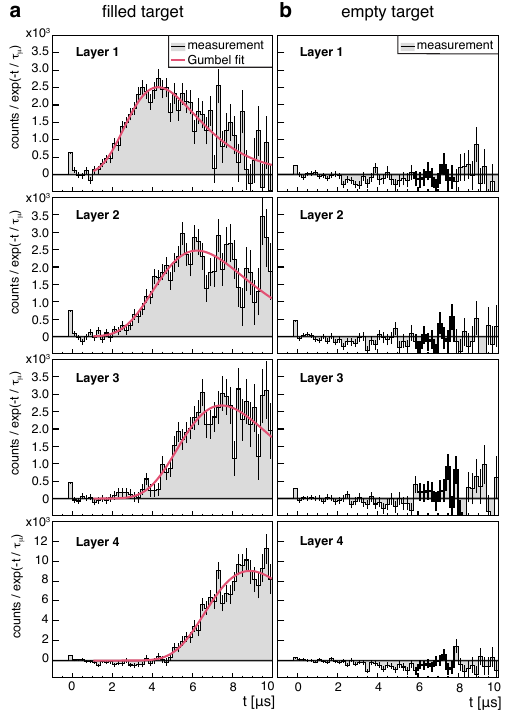}
\caption{\label{fig:measured timespec} \textbf{Time spectra in the four horizontal detector layers} Lifetime compensated spectra with background subtraction, when (a) He II is filled, and (b) in an empty target. In (a), a forward shifting peak is observed in the consecutive detector layers, implying \mm{} propagation. A Gumbel test function was fit to obtain the center of each peak, used to determine the mean propagation velocity of the Mu atoms. The error bars are statistical uncertainties of the measured counts prior to lifetime correction, scaled by $e^{t/\tau_\mu}$.}
\end{figure}

To overcome these limitations, the LEMING collaboration proposed a novel method to convert high-intensity $\mu^+$ beams to vacuum \mm{} using superfluid helium (He~II) \cite{soter_development_2021}, based on earlier muon moderation ideas \cite{taqqu_ultraslow_2011}, and the microscopic behavior of other exotic atoms in He~II \cite{soter_highresolution_2022}. Superfluid helium is known to be highly inert, repelling most impurities from its bulk. \mm{} as a light hydrogen isotope was also expected to have a large net positive chemical potential $E_{\rm c}$ when  placed in He~II \cite{taqqu_ultraslow_2011, saarela_hydrogen_1993, reynolds_hyperfine_1991}. If other interactions can be neglected, \mm{} atoms that reach the liquid surface would be hence ejected with a uniform velocity $|\vec{v}|=\sqrt{2E_{\rm c}/m_{\rm Mu}}$ along the surface normal, broadened only by the diffusion velocity prior to emission. If He~II is cooled below $T<0.3$~K, the saturated vapor satisfies ultra-high vacuum conditions, and scattering of the ejected \mm{} atoms becomes negligible for interferometry and precision spectroscopy \cite{antognini_roomtemperature_2022, soter_development_2021}. 

This concept relied on several unknown factors, like the exact value of $E_{\rm c}$,  the \mm{} formation probability below $0.7$~K in He~II \cite{abela_muonium_1993,krasnoperov_muonium_1992,eshchenko_excess_2002}, and most critically, the \mm{} thermalisation and diffusion process. Mu atoms are created when muons come to rest in matter, in average ranges of $h > \SI{30}{\um}$ below the He~II surface for conventional muon beams ($p_{\mu^+}> 12$~MeV/c). For efficient vacuum \mm{} conversion the average diffusion times $\tau_{\rm diff}$ should not exceed the $\mu^+$ lifetime. While this would not be possible from conventional bulk matter at $T<0.3$~K, in He~II the mobilities are only limited by scattering on the sparse thermal phonons, and complex hydrodynamic effects like vortex nucleation for large impurities \cite{mcclintock_ions_1984}. Theoretical predictions indicated that hydrogen and $^3$He isotopes \cite{saarela_hydrogen_1993} and exotic atoms like antiprotonic helium \cite{soter_highresolution_2022} cause small defects in He~II. If \mm{} could be modeled as a similarly small impurity, a rapid (near ballistic) diffusion could ensue without hydrodynamic drag, which was observed in the special case of $^3$He \cite{lamoreaux_measurement_2002}.

To demonstrate \mm{} emission to vacuum from He II, the experimental setup shown in Fig.~\ref{fig:Setup} was constructed at the Paul Scherrer Institute (PSI). A $\mu^+$ beam with momentum of $p_\mu = 12.5$ MeV/$c$ was bent downwards by a 30$^\circ$ angle (see Methods). The beam traversed a \SI{22}{\um}-thin scintillator foil that tagged the $\mu^+$ before entering the cryogenic chamber via a thin (\SI{6}{\um}) titanium window. The chamber was thermalized to $T \approx  \SI{0.2}{\K}$ by a dilution refrigerator and filled with a 2~mm layer of isotopically purified He II. The $\mu^+$ beam momentum was tuned to optimise stopping close to the He II surface, in an average stopping depth of $h\approx\SI{35}{\um}$ (Fig.~\ref{fig:Setup}b). While the $\mu^+$ at rest cannot capture an electron from ground state $^4$He atoms at these temperatures, it could recombine with the last free electron from its ionization trail.  We verified efficient \mm{} formation for the first time below 0.7~K with the observation of the \mm{} triplet spin precession (see Methods), extending the range of earlier measurements \cite{abela_muonium_1993}.


Mu emission from the liquid surface and propagation in vacuum was inferred by tracking positrons ($e^+$) from muon decays ($\mu^+\rightarrow e^+\nu_{e}\bar{\nu}_{\mu}$). For this purpose, a vertically segmented array of 16 plastic scintillators bars with silicon photomultiplier readout \cite{zhang_scintillation_2022} was mounted on the two sides of the chamber (see Fig.~\ref{fig:Setup}a and c). Coincidence conditions between detector segments defined geometric acceptance regions for the origin of decay positrons, as shown in Fig.~\ref{fig:Setup}c. Horizontal coincidences such as $\rm LC1 \land LF1$  (red shaded regions) measured the time of \mm{} decays in increasing distances from the He~II surface, while others like $\rm RC1 \land RF4$ (blue shaded area) monitored decays at the surface.

The resulting time spectra of $\mu^+$ decays in the four horizontal layers are shown in Fig.~\ref{fig:measured timespec}, where the time $t = 0$ was determined by the $\mu^+$ arrival on the entrance detector. In these histograms, the left and right coincidences were summed, and the counts and errors in each time-bin centered around $t$ were multiplied by $\exp (t/\tau_\mu)$ to compensate for the effect of $\mu^+$ decays. Decays of $\mu^+$ in the target walls and floors appeared as a constant background which was subtracted (see Methods). Any deviations in the resulting time spectra indicate a dynamic behavior, namely \mm{} propagating through the acceptance regions. With He II present in the target (Fig.~\ref{fig:measured timespec}a), the corrected time spectrum of Layer 1 showed a clear rise between $t=\SI{2}{\us}$ and $\SI{4}{\us}$, followed by a decline - indicating that Mu entered and exited the Layer 1 acceptance region above the liquid. In Layers 2-4, similar peak structures appeared at later times, indicating that Mu propagated upwards through the detection zones. In contrast, when the chamber was empty and the $\mu^+$ stopped in the copper floor (Fig.~\ref{fig:measured timespec}b), a consistently flat spectrum indicated no Mu emission. The time spectra in Fig.~\ref{fig:measured timespec}a were first fit with a generic skewed Gumbel distribution to obtain a model-independent determination of the main characteristics of the emission. A linear fit of the peak centers versus the height of each layer above the He~II  surface yielded a preliminary estimate of the \mm{} propagation velocity of $v\approx2.1~$km/s in vacuum and a diffusion time of $\tau_{\rm diff}\approx \SI{2}{\us}$ in He~II (see Methods).

\begin{figure*}
\includegraphics{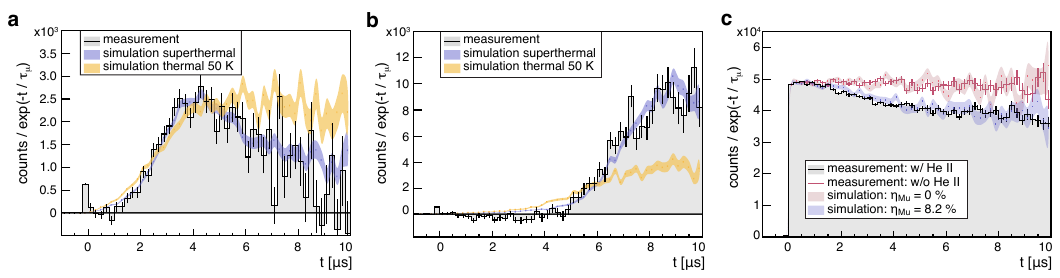}
\caption{\label{fig:MC simulation} \textbf{Comparison of measured and simulated time spectra.} Examples of three measured time spectra (solid histograms) superimposed with simulated ones (coloured bands) with $\pm 1 \sigma$ statistical deviations. Lifetime- and background compensated histograms show the \mm{} decays sampled with (a) C1$\land$F1  (Layer 1) coincidences close to the surface, and (b) far from it, with C4$\land$F4 (Layer 4) coincidence conditions. The simulated superthermal beam (blue) fits both spectra well, while even the best-fitting thermal beam (yellow) visibly deviates. (c) Lifetime compensated decays originating from the target bottom (C1$\land$F4 coincidence conditions) when filled with He~II (black) and empty (red), superimposed with the corresponding simulations showing the escape of \mm{} from He~II.}
\end{figure*}

For a detailed analysis, the measured time spectra were compared to Monte Carlo simulations \cite{agostinelli_geant4a_2003, roberts_g4beamline_2007}. The above preliminary estimate on $\tau_{\rm diff}$ is compatible with a fast, near ballistic propagation model, without complex hydrodynamic effects that hinder the mobility of large impurities \cite{mcclintock_ions_1984}. The effective diffusion velocity $v_{\rm diff}=h/\tau_{\rm diff}$ is hard to predict at $T<0.3$~K due to the unknown thermalization and diffusion process. As discussed in Methods, we simply assumed the existence of a limit velocity below which the ballistic propagation would take place. The \mm{} velocities in the liquid were hence sampled from flat distribution $[0,2\cdot {v_{\rm diff}}]$, with an isotropic direction from the formation point, defined by the stopping $\mu^+$. The average effective velocity ${v_{\rm diff}}$ was varied in the simulation. The effect of the chemical potential at the surface was simulated by adding a uniform velocity $\vec{v}$ to the diffusion velocity in the direction of the surface normal, where the norm $|\vec{v}|=v$ was also varied in the simulation. A separate model assuming a thermal source with a Maxwell-Boltzmann velocity distribution and a $\sim \rm cos\theta$ angular distribution was also considered, with the temperature $T$ varied in the simulation to find the best fit.

Figure \ref{fig:MC simulation} shows the measured time spectra (black histogram) alongside the two simulated beam models for Layers 1 (a) and 4 (b), with the best fitting parameters. A combined $\chi^2$ across all four horizontal layers quantified the goodness of the simulation under given parameters \cite{gagunashvili_chisquare_2010}. The best fitting thermal beam model (yellow band) was found at $T=50~$K, but it still failed to give a good reproduction of the data (reduced $\chi^2$ = 1.98, ndf = 459), while the best fitting superthermal beam model (blue) closely matched it (reduced $\chi^2$ = 1.03, ndf = 459). These results were obtained by a parameter sweep that yielded a longitudinal propagation velocity of $v = (2180^{+160}_{-130})$~m/s and an effective diffusion velocity of ${v_\mathrm{diff}} = (26^{+6}_{-4})$~m/s, with statistical uncertainties corresponding to a 90\% confidence region (see Methods). Assuming the longitudinal velocity originates purely from the chemical potential we obtained $E_c\approx 2.8~$meV. The above results were validated using an independent data set with a lower liquid level, yielding to consistent results. The uncertainties were driven mainly by statistics and the detector resolution, and were largely insensitive to changes in transverse velocities due to the one-dimensional tracking geometry. We assessed the sensitivity of the simulated time spectra to experimental uncertainties, such as initial $\mu^+$ beam parameters, degrader thicknesses and detector positions. In case of the mean diffusion velocity $v_{\rm diff}$ this contributed to an additional $\delta {v_\mathrm{diff}} \sim \SI{15}{m/s}$ systematic uncertainty, while the uncertainties of the propagation velocity $v$ remained unaffected.

\begin{figure*}
\centering
\includegraphics{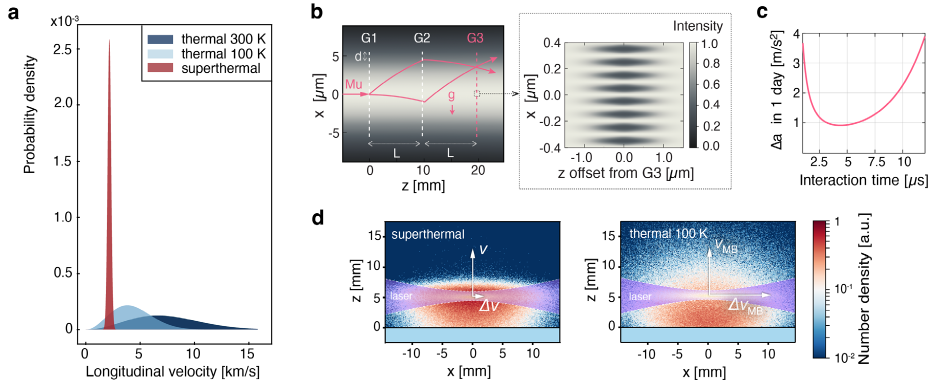}%
\caption{\label{fig:beam comparison} \textbf{Beam characteristics of the superthermal beam and possible applications.} (a) Velocity distribution of the \mm{} in vacuum, using the most pessimistic estimate ($\Delta v=150~$m/s) for the new superthermal source (red), thermal sources at room temperature (dark blue), and the coldest demonstrated (low intensity) source at 100 K (light blue). (b) Simulated intensity profile of the superthermal beam traversing the interferometer (grayscale gradient), with the schematic of the optimised grating positions. A zoom into G3 (right) shows the interference pattern. (c) Sensitivity to gravitational acceleration in 1 day at the $\pi$E5 beamline of PSI, assuming $d=100$~nm grating pitch, as a function of interaction time $t_i$. (d) Simulated \mm{} distributions, \SI{3.5}{\us} after the \mm{} formation, overlapping with a laser beam. The superthermal beam and a thermal source at 100 K are compared assuming the same diffusion time.}
\end{figure*}

The diffusion velocity in He~II provided a lower bound for the velocity spread ($\Delta v\geq v_{\rm diff}$) and the angular divergence ($\alpha\geq v_{\rm diff}/v\approx 12$~mrad) of the emitted beam. Interactions at the surface (like scattering on impurities, or surface waves) may result in additional broadening of the vacuum velocity distribution, which we investigated in simulations by comparing the rising edges of the time spectra. An isotropic velocity spread in the range of $(80<\Delta v<150)$~m/s was found to be consistent with the data (see Methods). Even using the conservative upper limit for $\Delta v$ ($\approx150$~m/s) 
(Fig.~\ref{fig:beam comparison}a), the new beam is clearly superthermal ($\Delta v \ll v$).

The fraction of emitted \mm{} atoms with respect to the stopped muons $\eta_\mathrm{Mu}={\rm\mm_{vac}}/{\mu^+_{\rm stop}}$ was determined from monitoring decays in the target using the C1$\land$F4 coincidences (Fig. \ref{fig:MC simulation}c). The decrease of the lifetime corrected counts indicated \mm{} leaving the acceptance area - the vicinity of He~II - and decaying elsewhere. Using the above determined velocities for the diffusion and emission process, $\eta_\mathrm{Mu}$ was varied as a parameter in the simulations, and yielded  $\eta_\mathrm{Mu} = (8.2^{+0.8}_{-1.1}) ~ \%$ (reduced $\chi^2$ = 1.13, ndf = 99). 
This value is consistent with a $\sim$ 70\% \mm{} formation efficiency, and consequent lifetime losses during diffusion. 

\begin{table}[b]
\caption{\label{Tab: Muonium yield} Comparison of potential vacuum Mu rates $R_\mathrm{Mu} = \eta_\mathrm{Mu} * R_\mu$ of various sources, at the optimized $\mu^+$ momenta ($p_\mu$) for each source. The initial muon rates $R_\mu$ for aerogel and He~II (SiO$_2$ film) targets are based on the yields of the $\pi$E5 (LEM) beamline at PSI, collimated to a $10\times10$~mm$^2$ aperture.}
\begin{tabular}{@{}lcccc@{}}
\midrule \midrule
Target & $p_\mu$ [MeV/$c$] & $R_\mu$ {[}s$^{-1}${]}  & $\eta_\mathrm{Mu}$ [\%] & $R_\mathrm{Mu}$ [s$^{-1}$] \\ \midrule 
Aerogel       & 23       & \SI{1.1e7}{}  & 3.5   & \SI{4e5}{}    \\
SiO$_2$ film  & 1        & \SI{3.0e3}{}    & 20    & \SI{6e2}{}      \\
He II         & 12.5     & \SI{1.2e6}{}  & 8.2   & \SI{1e5}{}    \\ \midrule \midrule
\end{tabular}
\end{table}

Using the muon rates $R_\mu(p_{\mu})$ of the highest intensity ($\pi$E5) beamline of PSI at a momentum used in the present study ($p_\mu=12.5~$MeV/$c$), the projected \mm{} rates can be estimated using the above conversion efficiency $R_\mathrm{Mu} = \eta_\mathrm{Mu} \cdot R_\mu$. The superthermal source in comparison with the projected performance of existing vacuum Mu sources are summarized in Table \ref{Tab: Muonium yield} (see Methods). The He II target yields vacuum \mm{} rates comparable to the best-performing room temperature aerogel targets \cite{zhao_optimization_2024}, while its longitudinal velocity distribution shown in Fig.~\ref{fig:beam comparison}a is an order of magnitude more narrow than the coldest (low yield) thermal sources at 100~K \cite{antognini_muonium_2012}. 


The feasibility of a \mm{} gravity experiment depends strongly on the parameters of the new beam. Considering a generic three-grating interferometer sketched in Fig.~\ref{fig:beam comparison}b, with $L$ distance between the gratings, diffraction on the first two gratings (G1, G2) creates interference fringes at the position of the third (G3). Gravitational acceleration shifts this pattern vertically $\Delta x=g_{\rm Mu}t_i^2/2 $, where $t_{\rm i}=L/v$ is the time of flight between the gratings. This phase shift can be measured by counting the transmitted \mm{} atoms as a function of the vertical displacement of G3 \cite{soter_development_2021}. The smallest observable acceleration ($\Delta a$) can be expressed as \cite{oberthaler_inertial_1996}: 
\begin{equation}
\label{Eq: Sensitivity}
     \Delta a = \frac{1}{2\pi t_{\rm i}^2}\frac{d}{C \sqrt{N}}
\end{equation}
where $C$ is the contrast of the interference pattern, $d\approx100$~nm is the grating period chosen from fabrication constraints \cite{savas_achromatic_1995}, and $N$ stands for the total number of detected atoms after G3. Starting from $N_0$ atoms, $\mu^+$ decays limit the detected number as $N\propto N_0e^{-2t_{\rm i}/\tau_\mu}$, which defines an optimum interaction time for the highest sensitivity $t_i\approx \SI{4.5}{\us}$ (Fig.~\ref{fig:beam comparison}c), and the corresponding grating spacing $L=9.6~$mm. 

Conservative calculations based on partially coherent waves \cite{mcmorran_model_2008} predict a high fringe pattern with a contrast $C > 0.35$ over an extended $\sim$\SI{1}{\um} region along the optical axis, which is amenable for grating alignment (see Methods). The small angular divergence also minimizes aperture related losses, which would be prominent using thermal beams. Combined with the expected Mu rate $R_\mathrm{Mu} \sim \SI{1e5}{s^{-1}}$ at PSI, this source would to enable a measurement of the gravitational acceleration with $\Delta a/g\approx0.01$ relative precision within $\sim$100 days of measurement.

In prospects of Mu 1S-2S laser spectroscopy, a high number density of \mm{} atoms within the laser beam waist diameter $\omega< 1$~mm is critical. Simulated spatial distribution of \mm{} atoms 3.5 \si{\us} after the muon arrival (Fig. \ref{fig:beam comparison}d) show the superthermal beam and a 100~K thermal source from porous SiO$_2$ proposed for the latest spectroscopy experiments \cite{crivelli_mumass_2018}, using the same diffusion time for fair comparison. While a thermal source is rapidly dispersing, the superthermal beam forms a compact, non-dispersive cloud only broadened by the diffusion time differences. Scaling with the rates in Table \ref{Tab: Muonium yield}, nearly $10^3$ times more \mm{} atoms could be addressed by the laser using the superthermal source. The lower velocity of the atoms would also decrease transit-time broadening ($\propto v/\omega$) by a factor two, while the narrow velocity distribution would reduce the main uncertainty in second-order Doppler shift corrections ($\propto v\Delta v$) by an order of magnitude. These advantages, combined with state of the art excitation schemes \cite{crivelli_mumass_2018}, could push the precision of the 1S–2S transition frequency measurement to the sub-kHz level. 

In this work we demonstrated a novel method to efficiently convert conventional $\mu^+$ beams to vacuum \mm{} in superfluid helium, by using the chemical potential to propel the atoms to vacuum. The narrow velocity distribution ($\Delta v<150~$m/s) against the longitudinal velocity ($v\approx2180$~m/s) makes this source superthermal, similar to high quality atomic beams from a nozzle, while yields were found comparable to the best diffuse thermal \mm{} sources. Preliminary estimates showed that the new beam is amenable to interferometry, and a subsequent percent-level precision in testing WEP in the second generation of the SM. The high rates and increased beam quality also provide a paradigm shift in laser spectroscopy experiments, allowing precise determination of the muon mass and tests of bound-state QED. The present efforts of the LEMING collaboration are directed to exploit these results, among them to demonstrate the first Mu interferometry.

\section*{Methods}

\subsection*{Design of the beamline and cryogenic converter}

\begin{figure*}[h]
\centering
\includegraphics[width=0.7\textwidth]{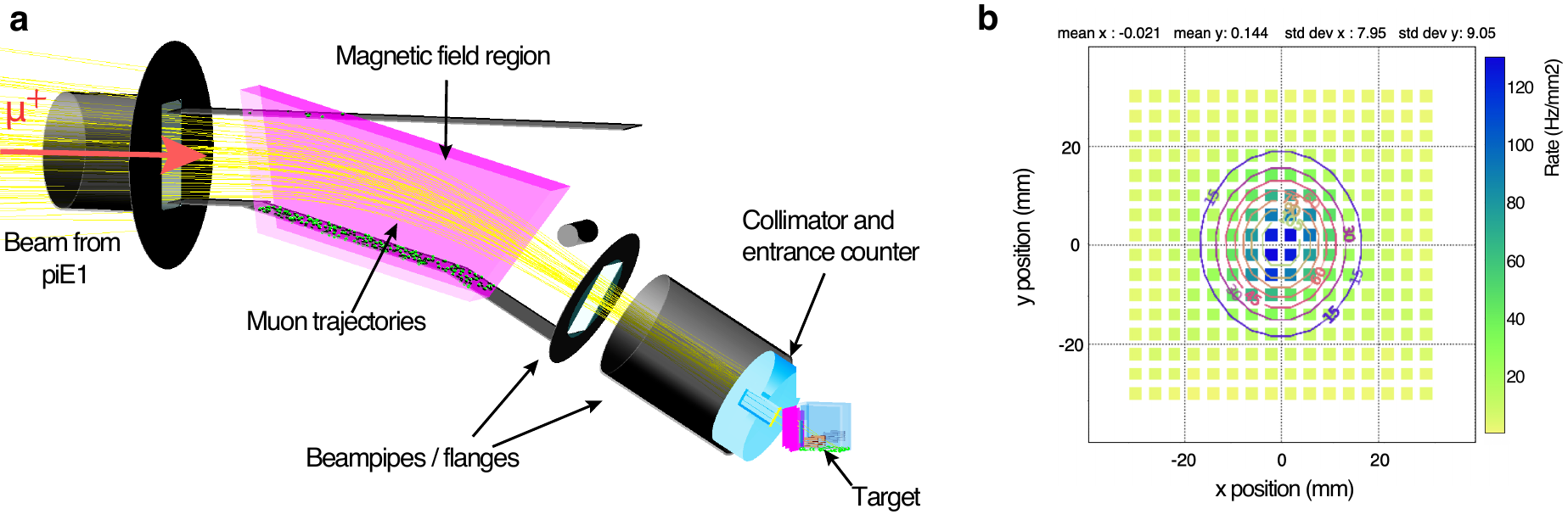}
\caption{\label{fig:beamline} \textbf{Simulation of the new beamline segment and the measured beam profile} (a) Monte-Carlo simulations of the  muon beamline, using magnetic field maps from finite element simulations. (b) Measured beam profile at the position of the entrance counter, using a retractable beam scanner with scintillator tiles and silicon photomultipliers.
}
\end{figure*}

Positive muons from the low intensity $\pi$E1 beamline \cite{PiE1_2024} of the CHRISP facility of PSI were delivered using a dedicated beamline segment with a $30^{\circ}$ vertical downwards bend. The magnetic elements and beam propagation was optimized using COMSOL multiphysics \cite{comsol} and Geant4 packages \cite{agostinelli_geant4a_2003, roberts_g4beamline_2007} (see Fig.~\ref{fig:beamline}a). The vertical bending magnet designed for this purpose had $12\times6$ turns of laminated coils carrying up to 60~A current, and a H-type yoke that defined a trapezoid magnetic field. The aperture width of $\sim80~$mm and average field length of 40~cm was a compromise between magnetic field quality and particle transmission. The collective focusing power of the last $\pi$E1 quadrupole and the trapezoid dipole field resulted in a nearly symmetric beam profile. This profile was measured and optimized using a retractable beam scanner, resulting in a <$15~$mm FWHM beam waist, and a rate of $R_\mu \sim 1.5 \times 10^{4} \, \mu^+ /$s at $p_\mu$ = 12.5~MeV/$c$ (Fig.~\ref{fig:beamline}b). 

The $\mu^+$ beam was passing through a 40-mm-thick brass collimator placed in front of the cryogenic chamber, with a $12\times25~\rm mm^{2}$ vertical aperture. The muon entrance detector that tagged the incoming $\mu^+$ was mounted to the exit of this collimator. It was composed of a ($22 \pm 7$)~\si{\um} thin plastic scintillator foil (EJ-212), with the edges glued to an acrylic frame that efficiently coupled out the weak scintillation light to eight silicon photomultipliers (AdvanSiD SiPM). The tagged muons traversed three layers of thermal shielding with $\sim2~\si{\um}$ thick aluminized mylar foils covering apertures, before they entered the target chamber via the $(6.0\pm1.5)$\si{\um} titanium beam window. 

The central momentum $p_{\mu}$ and momentum bite $\Delta p_{\mu}/p_{\mu}\approx 1.5 \%$  (FWHM) was set by scaling magnet currents and adjusting slits in the beamline following predefined values for surface muon beams, which was prone to uncertainties. The stopping power carried additional uncertainties from material thicknesses of the Ti foil and the entrance detector described above. Experimental ranging of the $\mu^+$ beam was hence carried out by tuning the magnet currents and monitoring the number of $\mu^+$ decays in the target, where a sudden increase indicated the limit $p^0_{\mu}$ where the beam passed the degrader layers. The measured $p^0_{\mu}$ was $ 0.3$~MeV/c smaller compared to expectations from simulations, which is well within the above uncertainties. In further simulations, the corrected $p_{\mu}$ = 12.5~MeV/$c$ experimental momentum and the nominal degrader thicknesses were used to obtain the stopping distribution shown in Fig. \ref{fig:Setup}b resulting in the mean depth of $h=35$~\si{\um}. The ranging uncertainties yield an uncertainty of $\delta h \sim 15$~\si{\um} which was taken into account while estimating systematic uncertainties of the diffusion velocity.

The He~II target was contained in an oxygen-free high conductivity copper chamber mounted to a custom-built dilution cryostat  \cite{vandenbrandt_compact_1990} with tail and shielding optimized for this experiment. The side walls of the chamber were thinned down towards the detector array to $<1$~mm to minimize scattering of the decay positrons. The target gas was isotopically pure $^4$He  with an isotopic ratio of  ${\rm ^3He}/{\rm ^4He}\approx 10^{-12}$. The $^4$He gas was iteratively measured and filled from a room temperature buffer volume of \SI{1.0}{L} and delivered to the chamber by two capillaries. With this method the depth of the liquid surface could be adjusted with $\pm$ 0.3~mm precision.

The high isotopic purity was vital to observe the emitted \mm{} atoms. Our previous attempt using high purity (99.9999 \%, 6N) commercially available helium gas with natural abundance (${\rm ^3He}/{\rm ^4He}  \sim 10^{-6}$) did not result in vacuum emission, which was accounted for $^3$He impurities collecting at the surface and acting as scattering centers.

The temperature was measured with \SI{5}{mK} precision using two calibrated Cernox sensors from Lake Shore Cryotronics, attached to the top and bottom of the chamber. Since the exact temperature below $T<0.3~$K was less relevant for the observed phenomenon, the cryostat was allowed to run free at the base temperature, which drifted in time between $T_{\rm top}\approx170 - 195$~mK and $T_{\rm bot}\approx 225 - 235$~mK on the two sensors respectively. 

\subsection*{Observation of \mm{} formation at 200~mK}

\begin{figure}[h]
\centering
\includegraphics[width=\linewidth]{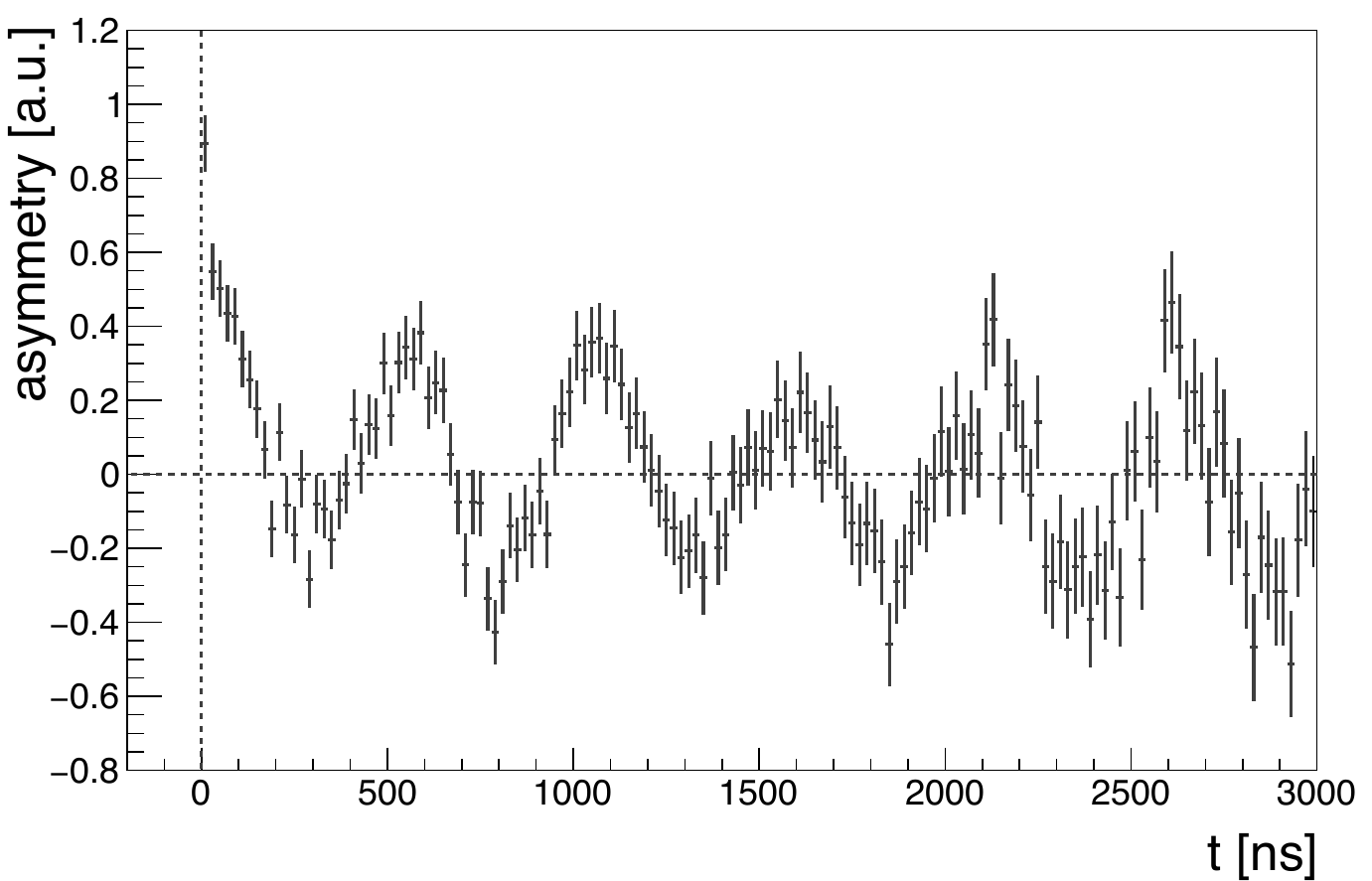}
\caption{\label{fig:Mu SR} \textbf{Muonium triplet precession in the target} Left-right asymmetry spectrum of C1$\land$F4 coincidences, monitoring spin-dependent decays inside superfluid helium. The Larmor precession signal indicates that $\mu^+$ stopped in He II formed Mu atoms, and the early onset of the amplitude implies formation on a $100~$ns timescale. 
}
\end{figure}

Muonium formation was observed by their triplet precession in the weak residual ($\sim50-100~\si{\uT}$) magnetic fields, dominated by the Earth magnetic field \cite{yaouanc_muon_2010}. Positive muons emit decay positrons preferentially in the direction of their spin, hence spin precession of initially polarized fraction of the muons or muonium atoms appear as an oscillatory signal in the surrounding detectors. This signal can be enhanced by looking at the time-dependent asymmetry $(N_\mathrm{l} - N_\mathrm{r}) / (N_\mathrm{l} + N_\mathrm{r})$ between left ($N_\mathrm{l}$) and right-sided detector counts ($N_\mathrm{r}$), using LC1$\land$LF4 and RC1$\land$RF4 coincidences as shown in Fig \ref{fig:Mu SR}. At such low fields the muonium triplet precession $\omega_{\rm Mu}$ is the only visible in \si{\us} timescales, being significantly faster than the muon precession, $\omega_{\mu}\approx \omega_{\rm \mm}/105$. The early appearance of the asymmetry implied that most of the \mm{} atoms formed promptly upon stopping, or on time scales $\tau_{\mu\rightarrow\rm\mm{}}\ll1/\omega_{\rm Mu}$, which agrees with earlier observations made at higher temperatures \cite{eshchenko_excess_2002}.

A quantitative muon-spin resonance measurement would have needed a well controlled magnetic field, but relative changes of the amplitudes were monitored and no significant change was seen below $0.7$~K. This indicated that formation efficiencies were not changing drastically at low temperatures from the previously measured $\sim 90\%$ at 0.7~K \cite{abela_muonium_1993}, which agreed with the $>70\%$ formation efficiency implied from the present emission measurements as well (see below).

\subsection*{Processing of the time spectra}
\begin{figure}[h]
\centering
\includegraphics[width=\linewidth]{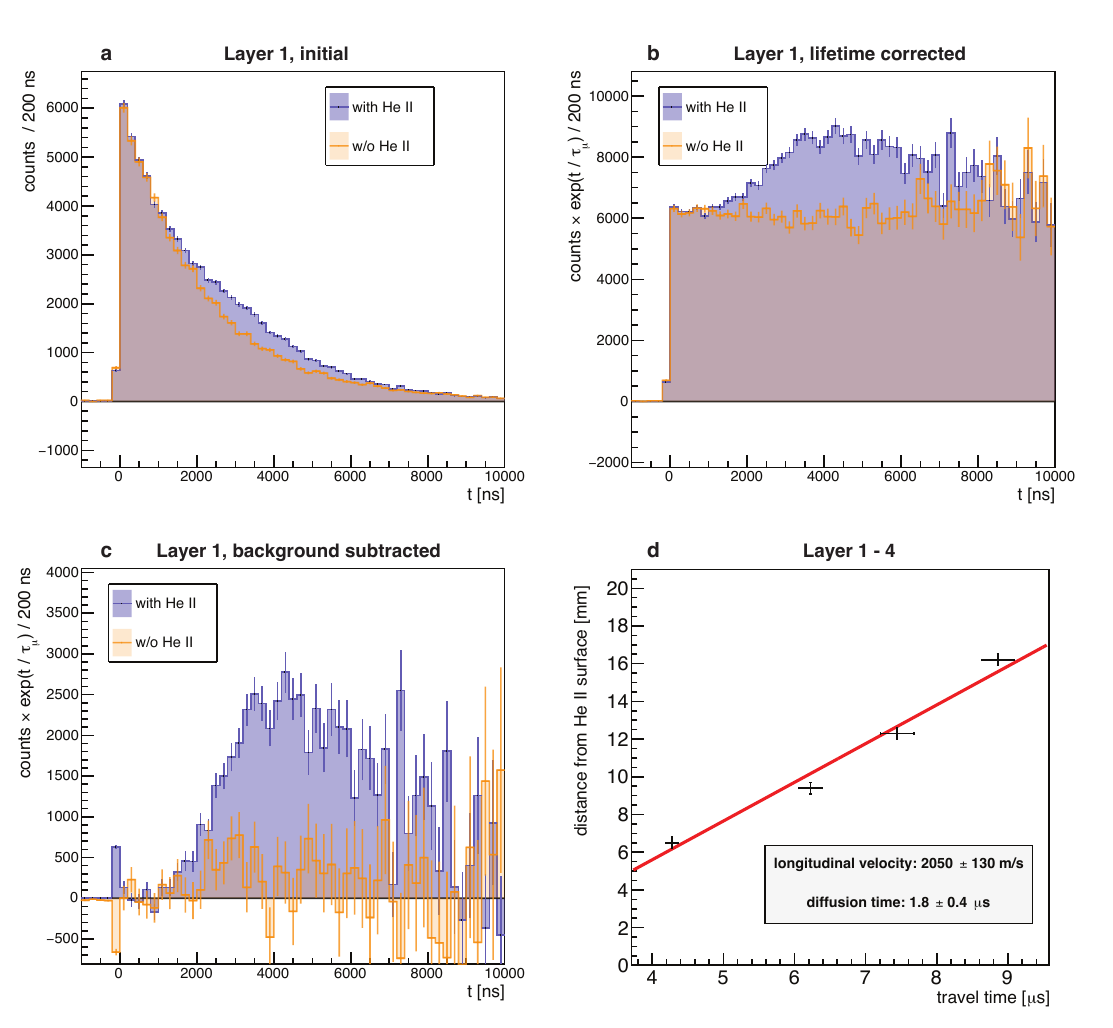}
\caption{\label{fig: Linear fit Gumbel} \textbf{Model independent time spectra analysis.} (a) Raw time spectrum (rebinned) from the digitizers $N_{\rm raw} (t)$, dominated by the exponential $\mu^+$ decay. (b) Lifetime-correct spectrum $N_{\rm lt} (t)$, yielding a flat background with a peak arising from Mu dynamics. (c) Background-subtracted spectrum $N_{\rm final} (t)$, where deviations from zero indicate Mu emission into vacuum. All spectra show the Layer 1 coincidences for a measurements with He II (blue) and an empty chamber (orange). (d) Linear fit of the detector distance from the He~II surface against the extracted mean arrival time used to determine the propagation velocity and diffusion time of \mm{}.}
\end{figure}

The time spectra shown in Fig. \ref{fig:measured timespec} were constructed from coincidence counts between a close and a far detector within a 40~ns coincidence window, while $t=0$ was defined by the entrance counter. Spectra of the left $N_\mathrm{l}(t)$ and right $N_\mathrm{r}(t)$ detector pairs were summed to increase the solid angle, and to cancel the effects of spin-precession: $N_\mathrm{raw}(t) = N_\mathrm{l}(t) + N_\mathrm{r}(t)$ (Fig. \ref{fig: Linear fit Gumbel}a). The counts and statistical error bars in each time-bin centered around time $t$ were multiplied by $\exp{(t/\tau_\mu)}$ to compensate for the muon decay: $N_\mathrm{lt}(t) = N_\mathrm{raw}(t) \cdot e^{t/\tau_\mu}$. In these lifetime-corrected spectra, the decays from $\mu^+$ in the target walls and floors appear as a constant background starting at $t=0$ (Fig. \ref{fig: Linear fit Gumbel}b). For Layers 1-4, the constant muon background was calculated by averaging the counts in the first \SI{1}{\us}, where no \mm{} atoms were expected within the detector acceptances. The background, $N_\mathrm{bg} = \frac{1}{N} \sum_{t = 0}^{t = 1 \si{\us}} N_\mathrm{lt}(t) $, with $N$ denoting the number of bins in the 0 to \SI{1}{\us} window, was subtracted from the lifetime-corrected spectra to obtain the spectra containing only the vacuum \mm{} signal $N_\mathrm{final}(t) = N_\mathrm{lt}(t) - N_\mathrm{bg}$ (Fig. \ref{fig: Linear fit Gumbel}c). The the histograms were rebinned only after this last step to 200~ns width. 

The resulting spectra were fit with a Gumbel distribution chosen as a skewed test function with minimum degree of freedom that describes the data well (reduced $\chi^2$ = 1.24, 1.26, 1.02, 1.57 for Layers 1-4 respectively, with ndf = 42). The fit function was $f(t) = f_0 e^{-(z+e^{-z})}$, with $z=\frac{t-\gamma}{\beta}$, and the fit parameters $f_0$, $\gamma$, and $\beta$ denoting the peak amplitude, peak position, and width respectively. The peak position $\gamma$ provided the mean travel time of \mm~ to each layer and is plotted against the height above the liquid surface in Fig. \ref{fig: Linear fit Gumbel}d. The slope of a linear fit yielded a preliminary estimate of the \mm{} propagation velocity in vacuum, $ \sim 2.1$~ km/s, while the time offset at zero distance from the liquid surface yielded a diffusion time in He~II ($\sim \SI{2}{\us}$). These values were used for initial parametrization of the Monte Carlo simulations, where we carried out a detailed analysis by varying them.

For the \mm{} emission plot using $\rm C1\land F4$ coincidences (Fig. \ref{fig:measured timespec}c), the background was not subtracted since dynamics may start at early times and the signal is an escape (as opposed to the appearance) of \mm{}. 

The number of escaping \mm{} atoms were also directly obtained from the original time spectrum by counting the missing decays in late time, under the exponential decay curve fitted to the first time bins ($t<\SI{1}{\us}$). This model-independent method yielded $\eta_{\rm Mu} = (8.6 \pm 0.8) \% $, consistent with results from the Monte Carlo simulations.

\subsection*{Simulation of \mm{} diffusion and emission}

The Mu beam emitted from He II was modeled using the Geant4 framework \cite{agostinelli_geant4a_2003, allison_recent_2016}. First, the stopping distribution of $\mu^+$ in He~II was simulated using experimental ranging as described above, and taken as the starting position of the \mm{} atoms. 

With the lack of a microscopic model, the thermalization and diffusion in He~II was modeled based on simple considerations and the measured transport properties of other impurities. The initially hot \mm{} atoms were assumed to thermalize first by creating collective excitations, until they reach the kinematic limit defined the dispersion relation, that features the roton minimum ($\Delta E_{\rm rot}/k_{\rm B}\approx 8.6~$K, $p_0\approx 1.9$~\AA{}$^{-1}$). For a heavy particle this limit would be the Landau velocity ($v_L\approx60 $~m/s), but the exact dynamics is sensitive on the effective mass of the impurity, and complex hydrodynamic effects. Most atoms and negative ions cause large (nm-sized) defects in He~II, and hydrodynamic backflow giving rise to large effective masses ($m^*>50~ m_{^4\rm He}$, with $m_{^4\rm He}$ the atomic mass of $^4$He) \cite{toennies_spectroscopy_1998, mcclintock_ions_1984}. For ions, the ensuing hydrodynamic effects like vortex nucleation limit the effective velocities in He~II to $v_{\rm max}<1~$m/s \cite{mcclintock_ions_1984} under saturated vapor pressure, which would not allow \mm{} atoms to escape He~II within their lifetime. Defects caused by \mm{} atoms were on the other hand expected to be much smaller, $m_{\rm \mm{}}^* \approx 2.5~m_{^4\rm He}$ based on the calculated and measured properties of hydrogen and helium isotopes \cite{saarela_hydrogen_1993,hayden1995atomic}. This value is closer to the effective mass of $^3$He in He~II $(m^*_{^3\rm He}\approx 1.5 m_{^4\rm He})$ which is known to exhibit fast (ballistic) diffusion, scattering only on thermally available phonons \cite{lamoreaux_measurement_2002}. The thermalization at $T<0.3$~K might also strongly depend on the relevant length scales, and the velocity distributions cannot be predicted. Instead of modeling it, we simply assumed the existence of an upper limit velocity for \mm{}, below which a near ballistic propagation takes place. The velocity of each \mm{} atom in He II was hence sampled from a flat distribution $[0,2\cdot v_{\rm diff}]$ and distributed isotropically from the point of origin defined by the $\mu^+$ stop.  Based on the scarcity of phonons at $T<200$~mK ($\rho_{\rm ph}\propto T^3$) \cite{lamoreaux_measurement_2002} and $^3$He impurities in the purified liquid, collisions during the observed \si{\us} diffusion times are unlikely, hence atoms were assumed to follow a ballistic trajectory to the surface. The average diffusion velocity of the ensemble  $v_{\rm diff}$ was a variable parameter in the simulation.

Mu emission into vacuum was modeled by adding an additional velocity component $\vec{v}$ to the diffusion velocity $\vec{v}_{\rm diff}$. Multiple scenarios were considered, with two presented in this work. (1) Superthermal beam: a constant $\vec{v}$ was added in the direction of the surface normal, motivated by the assumption that Mu gains a (mono-energetic) kinetic energy from the positive chemical potential upon leaving the flat surface of He II. (2) Thermal beam: we chose the Maxwell-Boltzmann velocity distribution that fit the overall experimental data the best ($T\approx50~$K), and directed it with a random sample from a $\cos \theta$ angular distribution, where $\theta$ is the emission angle with respect to the surface normal. The resulting Mu trajectories were propagated above the surface, and the detector response from subsequent decay positron trajectories were used to generate the simulated time spectra, which we compared to the measured ones.

\begin{figure}[h!]
\centering
\includegraphics[width=\linewidth]{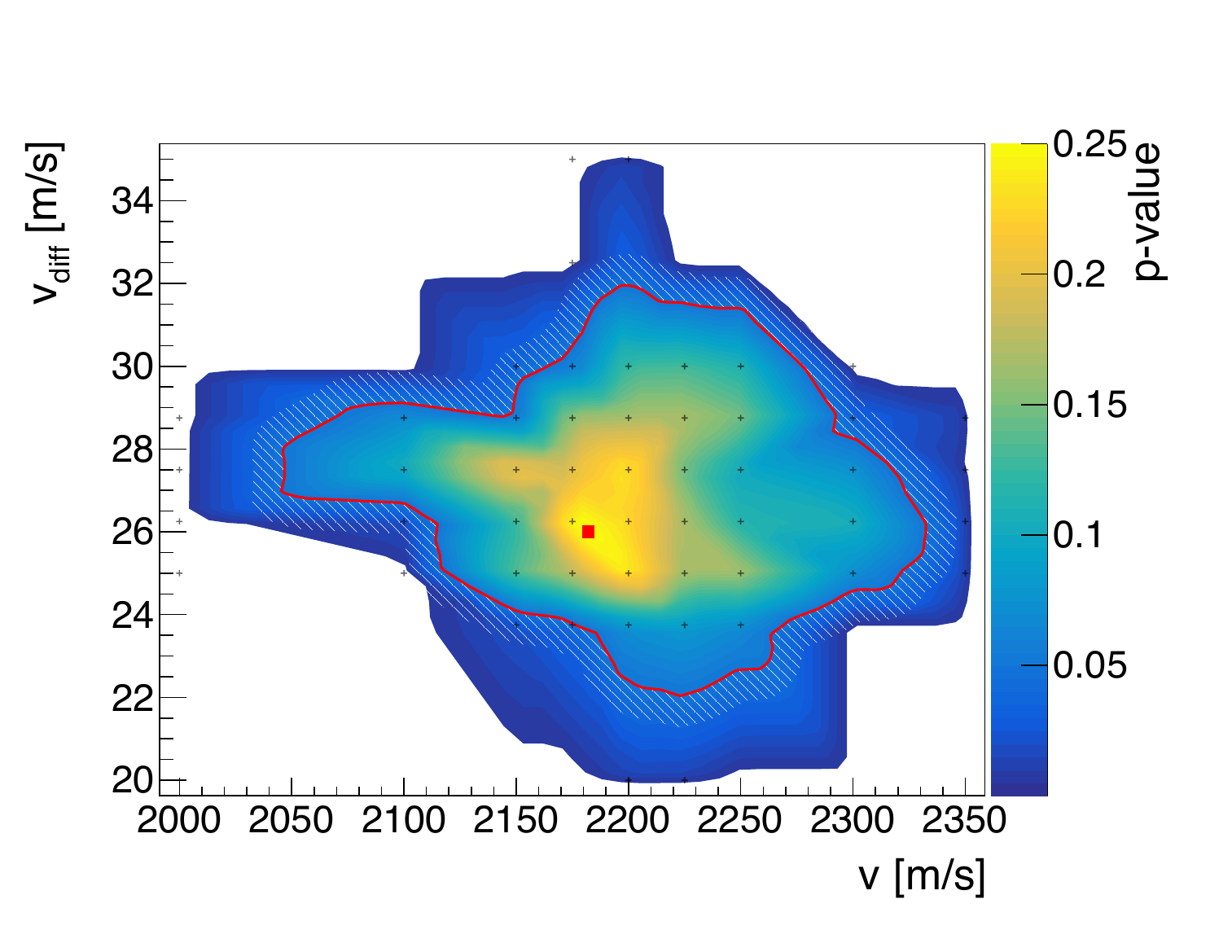}
\caption{\label{fig:Parameter sweep} \textbf{Parameter scan for the directed beam scenario.} The two parameters, ${v_{\rm diff}}$ and $v$, are varied (simulated values are indicated by black crosses), and the $p$-value from the $\chi^2$ test of homogeneity between measured and simulated time spectra are plotted. The red square marks the best fit and the red line shows the $p$-value = 0.05 boundary corresponding to a two-sided 90~\% confidence interval.}
\end{figure}

The directed beam model included two free parameters: the average diffusion velocity $v_{\rm diff}$ and the longitudinal velocity $v$ acquired upon surface ejection. We conducted a two-dimensional parameter scan, where for each pair the corresponding $p$-value from a $\chi^2$ test comparing simulated and measured time spectra was calculated \cite{gagunashvili_chisquare_2010}. The best-fit values correspond to the maximum $p$-value and we determined the uncertainties as the region where $p$-value = 0.05, representing a 90~\% confidence region. This yielded an average diffusion velocity of $v_{\rm diff}=(26^{+6}_{-4})$~m/s, and longitudinal velocity of $v = (2180^{+160}_{-130}$) m/s, as shown in Fig. \ref{fig:Parameter sweep}.

\subsection*{Estimation of velocity distribution beyond thermal contributions}
\begin{figure}[h]
\centering
\includegraphics[width=\linewidth]{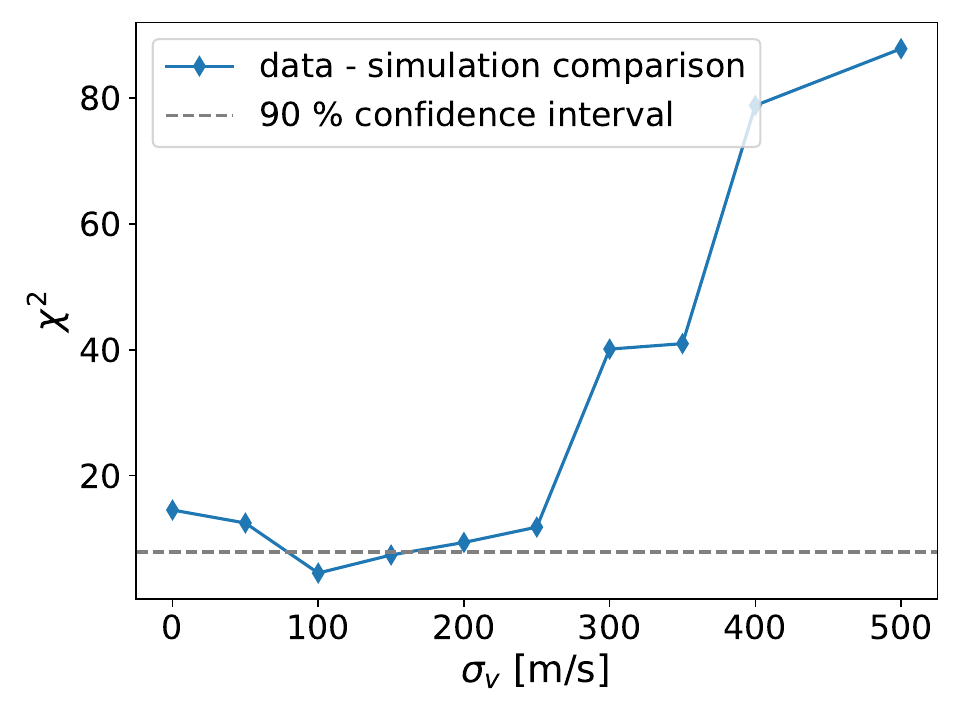}
\caption{\label{fig:Chi2 velocity dist} \textbf{Estimation of velocity distribution.} $\chi^2$ obtained from comparing the shape of the time spectra of simulated Mu beams with defined velocity broadening $\sigma_v$ and the measurement. Velocity spreads of 80 - 150 m/s are within a 90 \% confidence interval.}
\end{figure}

To asses the influence of possible surface interactions on the emergent Mu velocity distribution, independent simulations of Mu emission were performed, assuming for a generic description a Gaussian velocity profile centered at 2180 m/s with variable isotropic velocity spreads of $\sigma_v$. A large velocity distribution would cause the expansion of the cloud of atoms, and would appear as a change in the rising edge in the consecutive detector layers. For each $\sigma_v $, simulated time spectra were compared with the measured ones by calculating a $\chi^2$ from the steepness of the rising edges in Layer 2-4 relative to Layer 1. Velocity spreads in the range of $\sigma_v \approx$ 80 - 150 m/s were found still consistent with the data at the 90~ \% confidence level (see Fig. \ref{fig:Chi2 velocity dist}), with $\sigma_v \approx$ 100 m/s being the best fit. For evaluation of the beam quality, we chose the conservative upper limit $\sigma_v<150$~m/s, and assumed isotropic velocity spreads in all directions.

\subsection*{Sensitivity calculations for interferometry}

The preliminary estimate of sensitivity is based on simple considerations of a phase measurement in a 3-grating interferometer, see e.g. Ref.~\cite{oberthaler_inertial_1996}. Besides decay losses, the number of detectable atoms $N$ are further reduced by geometric factors, $N = N_0\alpha \vartheta_G^3 \eta_{D} \cdot e^{-2t_i/\tau_{\mu}}$. Here the acceptance ($\alpha$) of the interferometer accounts for the fraction of atoms that could pass the $1~{\rm cm}^2$ aperture of the first and third grating. Using a simple Monte Carlo model, thermal (Maxwell-Boltzmann) beams with $\sim\cos\theta$ angular distribution suffer large losses here, with $\alpha_{\rm th}\approx 0.07$ for the \SI{4.5}{\us} optimum time of flight length ($\overline{v}_{\rm th}\approx6600$~m/s, $L_{\rm th}\approx29~$mm). In contrast, the new superthermal beam is expected to have $\alpha\approx0.95$ acceptance. The transmission of a single grating ($\vartheta_G\approx 0.35$) and detection efficiency ($\eta_D \approx 0.75$) further decrease the measured $N$, where the above estimates are based on the geometry of the gratings and solid angle coverage of the detectors. 

The expected contrast ($C=\frac{I_{\rm max}-I_{\rm min}}{I_{\rm max}+I_{\rm min}}$) defined by the intensity minimum ($I_{\rm min}$) and maximum ($I_{\rm max}$) of the interference pattern was calculated using mutual intensity functions from partially coherent wave-optics, and a Gaussian Shell model of the particle beam~ \cite{mcmorran_model_2008}. The atoms have a de-Broglie wavelength of $\lambda_{\rm Mu}\approx 1.5$~nm, which results in a Talbot-length of $L_{\rm T}=\frac{2d^2}{\lambda_{\rm Mu}}\approx \SI{14}{\um}$. This would put the optimum grating spacing in the far field regime ($L\gg L_{\rm T}$). On the other hand, the large aperture size does not allow for the separation of the main diffraction orders at $L$ distance, meaning the second grating is still placed in the aperture near field.

A typical beam waist size of a focused muon beam defined the initial beam width as $A_0=10$~mm. The transverse coherence width $\ell_0$ of the new superthermal source was estimated conservatively from the upper limits of the angular divergence $\alpha\approx \Delta v/v\approx 70~$mrad, using the van Cittert-Zernike theorem $\ell_0\approx \lambda /\alpha = 22~$nm. The wavefront curvature $r_0$ of the superthermal source was assumed to be large compared to the length of the interferometer due to the flat surface of He~II. The resulting contrast depends also on the opened fraction $\nu$ of the grating, with $C=0.5$ for $\nu=0.45$. It should be noted that since the source is mainly incoherent ($\ell_0\approx d/4$), the real advantage of the superthermal beam is that the acceptable high contrast region is relatively wide ($> \SI{1}{\um}$) along the optical axis. This allows for grating misalignment (tip-tilt) in the longitudinal direction, leaving the rotational alignment of G2 and G3 the only critical point. Based on the small size ($L\approx 10~$mm) monolythic Si devices are considered. 

The above contrast calculation is conservative, as the effective aperture for this directed beam ($A_{\rm eff}< 1$~mm, region of G1 from where particles can reach G3) is significantly smaller than the source size, which should account as further collimation. Another clear advantage is the nearly mono-energetic nature of the beam, which predicts a nearly uniform interaction times and phaseshift under gravity.

\subsection*{Comparison of various Mu beams for laser spectroscopy}
To compare the suitability of different Mu sources for 1S-2S laser spectroscopy, we simulated Mu emission from three sources with their respective beam characteristics. The absolute vacuum Mu rate (see Tab. \ref{Tab: Muonium yield}) was calculated as $R_\mathrm{Mu} = \eta_\mathrm{Mu} \cdot R_\mu$, where the conversion efficiency $\eta_\mathrm{Mu}$ depends on the $\mu^+$ momentum. To estimate $R_\mu$, representing the maximal deliverable $\mu^+$ rate to a 10$\times$10~mm$^2$ target, we considered two beamlines at PSI. For subsurface muons, the $\pi$E5 beamline delivers $R_\mu =$ \SI{1.5e8}{} $\mu^+$/s at $p_\mu$ = 28 MeV/$c$ to a $\sigma_x \approx \sigma_y \approx \SI{10}{mm}$ beamspot \cite{dal_maso_optimization_2024}. Here, the muon rate follows a power law of $R_\mu \propto p_\mu^{3.5}$ \cite{pifer_high_1976}. For low-energy muons, the LEM beamline delivers $R_\mu =$ \SI{3e3}{} $\mu^+$/s at $E^\mathrm{kin}_\mu$ = 5~keV ($p_\mu$ = 1~MeV/$c$) to a $\sigma_x \approx \sigma_y \approx \SI{7}{mm}$ beam spot \cite{janka_improving_2024}. To estimate the number of atoms addressable by a cw laser, we defined $\phi_\mathrm{Mu}^\mathrm{laser}$ as the flux of Mu atoms propagating through two 1 $\times$ 1 $\si{mm^2}$ planes separated by 1 mm positioned at varying distances from the target surface. Compared to the proposed cryogenic SiO$_2$ source, the He II source increases the number of atoms that the laser addresses by a factor of $\approx 800$.

\section*{Data and Code availability}
The data sets and the analysis code will be made available upon request to the corresponding author.

\section*{Acknowledgement}
The authors would like to thank E. Krotschek for the discussion on the chemical potentials of hydrogen isotopes in superfluid helium. We are indebted to P. McClintock and D. Zmeev for providing the isotopically pure helium, and for having invaluable discussions about impurity mobilities. We thank the undergraduate students, R. Hamp and R. Koch for their enthusiastic support during the beamtime, the HIPA support groups at PSI for the stable beam, and the workshops at PSI and ETHZ for the precise and timely manufacturing of the mechanical parts. Last but not least, we are deeply saddened that we cannot thank Ben van der Brandt anymore, who provided invaluable guidance in cryogenic engineering for the early feasibility studies, and sadly could not see this result.

This work was funded by the SNSF Ambizione Grant No. PZ00P2 185975 and SNSF Project Fund 200441. 

\section*{Author contributions}

A.S. proposed the experimental method, based on ideas of D.T. for muon moderation, and earlier initiatives of K.K., A.A. and A.K.. The cryogenic target and connecting hardware was designed by A.S. The detector configuration was optimized by J.Z. using Monte Carlo simulations. The cryogenics detectors and electronics was developed by A.S., J.Z. and D.G.. The dilution unit and support for slow control was provided by M.B., and adapted by A.S., J.Z. and P.W. for this experiment. A.S., J.Z. and A.A. designed and built the bent beamline for muons. The thin foil entrance detector was designed by A.A. The retractable beam profile detector was developed by D.G. and A.S. The DAQ program was written by D.G., using frontend code and digitizers provided by F.W. and A.K. J.Z. carried out the Monte Carlo analysis, A.S. and R.W. the optical calculations. A.S. and J.Z. wrote the manuscript. All authors contributed to the data taking and the final formulation of this manuscript. 

\printbibliography

\section*{Extended Figures and Tables}


\end{document}